\documentclass[preprint,12pt]{elsarticle}

\usepackage{graphics}

\usepackage{amssymb}

\journal{}

\begin{document}

\begin{frontmatter}



\title{Magnetic Properties of Spinel FeCr$_{2}$S$_{4}$ in High Magnetic Field}


\author[kagoshima]{Masakazu Ito\corref{cor1}}
\ead{showa@sci.kagoshima-u.ac.jp}
\author[kagoshima]{Yuji Nagi}
\author[kagoshima]{Naotoshi Kado}
\author[kagoshima]{Shinpei Urakawa}
\author[kagoshima]{Takuro Ogawa}
\author[ISSP]{Akihiro Kondo}
\author[kagoshima]{Keiichi Koyama}
\author[tohoku]{Kazuo Watanabe}
\author[ISSP]{Koichi Kindo}
\cortext[cor1]{Corresponding author}
\address[kagoshima]{Department of Physics, Graduate School of Science and Engineering, Kagoshima University, Kagoshima 890-0065, Japan\\}
\address[tohoku]{Institute for Materials Research, Tohoku University, Sendai 980-8577, Japan\\}
\address[ISSP]{ISSP, University of Tokyo, Kashiwa, Chiba 277-8581, Japan\\}
\begin{abstract}
We studied temperature and magnetic field dependence of magnetization, $M(T)$ and $M(B)$, on thiospinel FeCr$_{2}$S$_{4}$.
We observed a step-like anomaly at $T_{OO} \sim$ 9 K for $M$($T$) attributed to an orbital order transition in magnetic field 0 $< B \leq$ 7 T. Furthermore, also for $M$($B$), an step-like anomaly appears at $B_{c} \sim$ 5.5 T below $T_{OO}$.   
This suggests an existence of a magnetic phase boundary at $B_{c}$ in $T < T_{OO}$. 
Because the anomaly at  $B_{c}$ of $M(B)$ exists in the orbital ordered phase, 
 the spin structure of FeCr$_{2}$S$_{4}$ in $T\ <\ T_{OO}$ is strongly coupled with the lattice of the system.
\end{abstract}


\end{frontmatter}


\section{Introduction}
 Spinel compounds described as the general formula $AB_{2}X_{4}$ attract much interests, because of showing various of physical properties.
The characteristic of spinel is that some of them have potential of technological applications.
Since colossal magnetoresistance ( CMR ) was reported on thiospinel FeCr$_{2}$S$_{4}$ and its derivatives\cite{Ramirez,Fitsch}, studies on FeCr$_{2}$S$_{4}$ are actively carried out. 
FeCr$_{2}$S$_{4}$ has a cubic spinel structure with a symmetry of $Fd\bar{3}m$. The Fe ions occupy at $A$ sites in a tetrahedral environment consisting of S$^{2-}$ ions at $X$ sites, and have valences Fe$^{2+}$.
The electronic configuration of Fe$^{2+}$ is 3$d^{6}$ $(\ e_{g}^{3}t_{2g}^{3}\ )$.
This suggests the $A$ site Fe$^{2+}$ ion is a John-Teller ( JT ) active ion.
Reflecting degree of freedom of $e_{g}$ orbitals, at $\sim$9 K, FeCr$_{2}$S$_{4}$ shows an orbital order transition\cite{Spender,Brossard, Feiner, Lotgering2}.
Meanwhile, Cr$^{3+}$ ions in this system occupy at octahedral $B$ sites.
The electronic configuration of Cr$^{3+}$ is 3$d^{3}$ $(\ t_{2g}^{3}\ )$, then Cr$^{3+}$ is a JT inactive ion.
\par
FeCr$_{2}$S$_{4}$ is a semiconductive ferrimagnetic compound with N$\acute{\rm e}$el temperature $T_{N} \sim 170$ K. 
 The magnetic interaction between the Cr$^{3+}$( spin angular moment $S = 3/2$ ) ions is dominated by ferromagnetic.  
The Fe$^{2+}$ ( $S = 2$ ) ions are strongly coupled with the Cr$^{3+}$ ions with an antiferromagnetic interaction.
From early studies of powder neutron diffraction\cite{Shirane, Broquetas}, ordered moment of the Fe and Cr ions were derived to be $4.2$ and $2.9\mu_{B}$, respectively. Therefore, $1.6\mu_{B}$ $(\ =\ $2Cr$^{3+}\ -\ $Fe$^{2+}\ )$ is expected as the saturated magnetic moment, assuming the collinear type ferrimangetic structure.
In a low magnetic field range, temperature $T$ dependence of magnetization $M(T)$ has a peak at around $T_{m} =$ 70 K, and $M(T)$ decreases with decreasing $T$ below $T_{m}$\cite{Tsurkan, Yang, Tong, Shen, Shen2, Tsurkan2}. An irreversibility of magnetization between zero-field-cooled (ZFC) and field-cooled (FC) procedures is observed below $T_{N}$ in a low $B$ range, and is reduced with increasing $B$\cite{Tong, Shen, Shen2}. 
An origin of this irreversibility has been explained by a domain wall motion with pinning centers by analyzing ac susceptibility and low-magnetic-field magnetization\cite{Yang, Tsurkan3}.
\par
 In this paper we report on high-magnetic-field magnetization of the polycrystalline FeCr$_{2}$S$_{4}$.
A new magnetic anomaly in field dependence of magnetization curve was observed.
\section{Experimental}
 The polycrystalline specimen of FeCr$_{2}$S$_{4}$ was prepared by a direct solid-state reaction.
High purity elements of Fe, Cr and S were mixed in stoichiometric ratio and reacted in a quartz tube at 950 C$^{\circ}$ for seven days.
The powder specimen was reground, pressed into a pellet and sintered at 950 C$^{\circ}$ for two days.    
Temperature $T$ dependence of dc magnetization $M(T)$ measurements were carried out using
a Quantum Design MPMS SQUID magnetometer.
Magnetic field $B$ dependence of magnetization $M(B)$ in 0 $\leq B \leq $ 18 T were measured by the hand made magnetometer with a extraction method. 
%
%
%
%
%
%
%
\section{Results}
\par
\subsection{M(T) of FeCr$_{2}$S$_{4}$}
\label{mt}
 Figure 1 shows a temperature $T$ dependence of dc magnetization $M(T)$ of FeCr$_{2}$S$_{4}$ at representative magnetic field $B$. 
With decreasing $T$, $M(T)$ increases rapidly around $T_{N}$ ( $\sim$ 170 K).
An irreversibility behavior between ZFC and FC process is observed in the low $B$ range, and with increasing $B$, irreversibility is suppressed as mentioned in the Introduction.  
Expanded plots of $M(T)$ curves in the low $T$ range are shown in Fig. 2 (a) and (b) for the range $ 0.05\leq B \leq0.2 $ T and $1.0\leq B\leq7.0$ T, respectively.
As was reported by previous studies\cite{Tsurkan, Shen2}, $M(T)$ shows a step-like anomaly in ZFC and FC process at around orbital ordering temperature $T_{OO}\sim$9 K. 
This anomaly comes from changing the orbital degree of freedom\cite{Tsurkan}, and indicates that the spin in this system is strong coupled with the lattice.
We determined $T_{OO}$ as a point at which $\partial M(T)/\partial T$ of FC process has a local minimum.  
In $B$  $\leq$ 0.1 T, $M(T)$ with FC and ZFC process show upturn and downturn behavior at $T_{OO}$, respectively. 
On the other hand, in $B$ $\geq$ 0.2 T, downturn of ZFC changes to upturn.  
  With increasing $B$, step-like anomaly is reduced and almost vanished at $B \sim$ 4 T.
Above $B =$ 5 T, however, step-like anomaly appears again at around $T_{OO}$ and this means that the orbital order phase remains even in high $B$ range.  
\subsection{M(B) of FeCr$_{2}$S$_{4}$}
\label{mb}
In order to investigate a detail of the step-like anomaly in $M(T)$, we measured isotherms $M(B)$ curves. 
Figure 3(a) and (b) show $M(B)$ curves in the range 0$ \leq B \leq$ 18 T at 4.2 and 20 K, respectively, as representations.
The $M(B)$ curves of both temperature show saturated behavior in $B > $ 12 T and reaches to the value of 1.6$\mu_B$. 
This value is close to the one expected the simple collinear ferrimagnetic structure, as mentioned in the Introduction.
The most pronounced feature is that, for $T =$ 4.2 K, $M(B)$ shows the small but non-negligible anomaly appears at around $B_{c} = $ 5.5 T as shown the arrow in the inset of Fig. 3(a). 
On the other hand, for $T =$ 20 K, $M(B)$ increases monotonically by applying $B$ and shows the no anomaly.  
\par 
Magnetic field derivative of magnetization, $\partial M(B)/\partial B$ is shown in Fig. 4 for various temperature.
This plot is very useful to detect the tiny magnetic anomaly such as seen in here.
 In the temperature range $T \leq$ 6 K, hump-like anomalies are observed and a position of that, $B_{c} = $ 5.5 T, is independent of temperature. 
 On the other hand, in $T \geq$ 10 K, no anomaly is observed. 
 This anomaly seems to exist just in the orbital ordered phase.  
\section{Discussion}
\subsection{Origin of the anomaly in M(B)}
The studies of structural distortion at $T_{OO}$ on FeCr$_{2}$S$_{4}$ are still carried on intensively.
Recently, broadening of diffraction lines at around $T_{OO}$ suggesting splitting of Bragg reflections was reported from high resolution x-ray synchrotron powder diffraction\cite{Tsurkan2}.
A reduction of the crystal symmetry should be related to magnetic properties of this system.  
 As a first possibility for the origin of the anomaly at $B_{c}$ in M(B), one can consider the domain wall motion with pinning centers as seen in $M(T)$ below $T_{OO}$.
Changing to the lower symmetry from cubic structure is likely to make additional pinning center.
In this picture, a large irreversibility gap between values of $M(B)$ in increasing and decreasing $B$ should be appeared below $B_{c}$.
In addition, the position of $B_{c}$ might be sensitive to $T$, because the domain wall is given a kinetic energy from thermal energy.
In our results, however, a large irreversibility are not observed and $B_{c}$ is very insensitive to temperature as shown in Fig. 3(a) and Fig. 4.  
Therefore, this picture is likely to ruled out.
\par 
As the second, the possibility of that magnetic field induced magnetic phase transition occurs at $B_{c}$.
Recent $\mu$SR investigation of FeCr$_{2}$S$_{4}$ predicted that the formation of incommensurately modulated noncollinear spin arrangement below $\sim$ 50 K\cite{Kalvius}, and a spin-reorientation due to this formation is cause of the cusp-like anomaly of $M(T)$ at $T_{m} \sim$ 70 K in the low $B$ range. 
Meanwhile, in the high $B$ range, $M(T)$ in $T < T_{m}$ is recovered gradually by applying $B$.
This may suggest that the non-collinear structure is quenched and collinear one is stable by applying $B$. 
In isotherm of $M(B)$, as shown in Fig. 3, the value of magnetic moment reaches to $\sim$1.6 $\mu_{B}$ at 18 T and this close to the one expected the simple collinear ferrimagnetic model, as mentioned in the Introduction.  
For $T > T_{OO}$, $M(B)$ shows gradual increase with increasing $B$ as shown in Fig. 3(b), and this  means that incommensurate-noncollinear spin arrangement changes to the commensurate-collinear one gradually. 
Meanwhile, for $T < T_{OO}$, $M(B)$ increases with the spin-flop or -flip like anomaly at 
$B_{c}$ and reaches to the saturation $\sim$1.6 $\mu_{B}$ as shown in Fig. 3(a).   
This means that, in the orbital ordered phase, the recovery of $M(B)$ from the incommensurate-noncollinear to commensurate-collinear spin structure is accompanied with the phase transition.
Since the anomaly at $B_{c}$ exists just in the orbital ordered phase, spin structure refer to the crystal structure  which has the lower symmetry than cubic as mentioned above.
\subsection{B-T phase diagram}
At last, we propose a magnetic $B-T$ phase diagram of FeCr$_{2}$S$_{4}$, which is summarized in Fig. 5(a) and (b) plotted with expanded a horizontal scale. 
The $B$ dependence of $T_{N}$ determined by $\partial M(T)/\partial T$ is also plotted in Fig. 5(a). With decreasing temperature, the system changes from paramagnetic ($P$) to ferrimagnetic ($I$) at around 170 K in $B <$ 7 T.  
\par
In the range $T_{OO} < T < T_{N}$, the magnetic phase $III$ is located in the range $B \geq B_{c}$, additional to the magnetic phase $I$ below $B_{c}$. 
The orbital order transition may occur simultaneously with the magnetic phase transition of $III$ at $T_{OO}$. 
It is reported that $C_{P}(T)$ of FeCr$_{2}$S$_{4}$ has a tail just above $T_{OO}$ in high magnetic field $\sim$10T\cite{Tsurkan, Shen}.
This might be referred to the phase transition from $III$ to $I$ phase.
To know the exact magnetic structure, the microscopical studies including the crystal structure in high magnetic field are needed. 
\section{Conclusion}
We have studied the magnetic properties in the high magnetic field on the spinel compounds FeCr$_{2}$S$_{4}$.
Irreversibility between field cooled and zero field cooled magnetization $M(T)$ is observed in the low magnetic field range. 
From temperature and magnetic field dependence of magnetization, $M(T)$ and $M(B)$, we proposed the existence of the new magnetic phase, below $\sim$9 K and above $\sim$5.5 T.  
Assuming the incommensurate-noncollinear spin structure in $T <$ 60 K predicted by recent $\rm{\mu}$SR work\cite{Kalvius}, high magnetic field gives rise to stabilize for the commensurate-collinear spin structure.
Because new phase boundary exists in orbital-ordered phase, the spin structure of FeCr$_{2}$S$_{4}$ in $T\ <\ T_{OO}$ is strongly coupled with the lattice. 
\\
\\
$\textbf{Acknowledgements}$
\\
\par
This work was supported by KAKENHI(Grant No. 23540393) from Japan Society for the Promotion of Science(JSPS).
\begin{figure}
\includegraphics[height=70mm,clip]{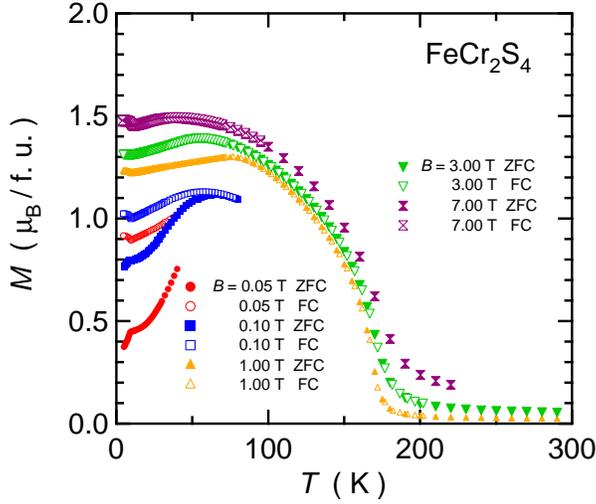}
\caption{
Temperature $T$ dependence of dc magnetization $M(T)$ of FeCr$_{2}$S$_{4}$ at representative magnetic fields. 
 }\label{Fig1}
\end{figure}
\begin{figure}
\begin{center}
\includegraphics[height=140mm,clip]{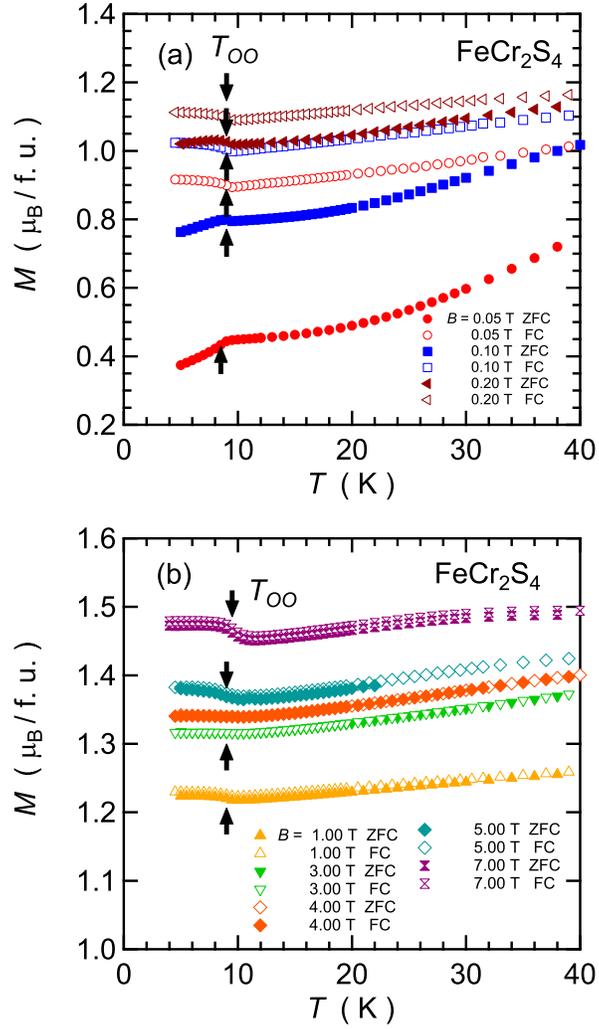}
\caption{
Expanded plot of $M(T)$ in the low $T$ and in the range (a) 0.05 $\leq B \leq$ 0.2 T and  (b) 1.0 $\leq B \leq$ 3.0 T. The arrows show the position of $T_{OO}$ determined by $\partial M(T)/\partial T$.   
}
\label{Fig2}
\end{center}
\end{figure}
\begin{figure}
\begin{center}
\includegraphics[height=140mm,clip]{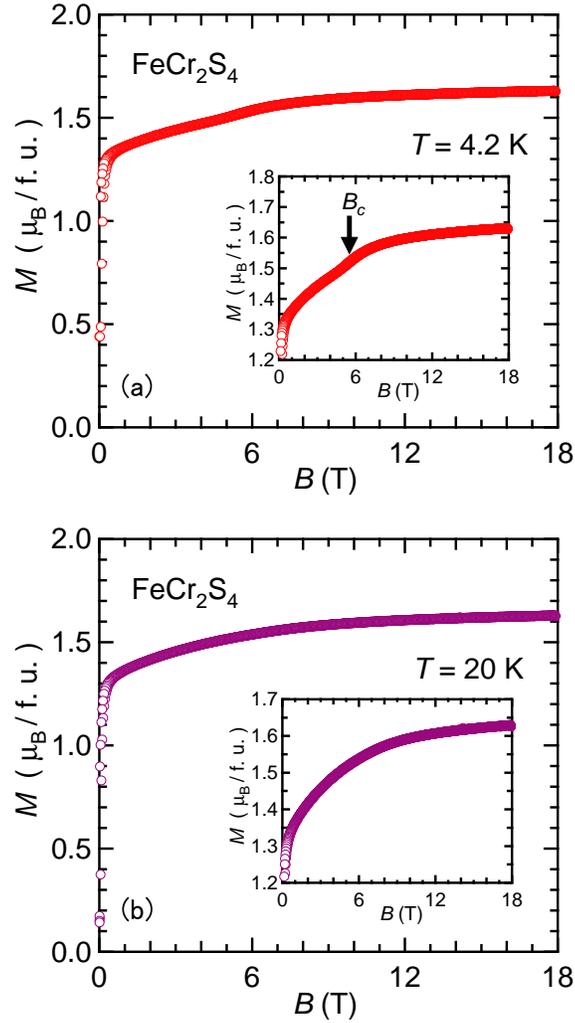}
\caption{
Magnetization $M$ of FeCr$_{2}$S$_{4}$ as the function of the magnetic field $B$, $M(B)$, in the range 0 $\leq B \leq$ 18 T at (a) $T =$ 4.2 K and (b) $T =$ 20 K. 
The insets of both panels are plots with expanded vertical scale.
The arrow in the inset of (a) shows the temperature position of the step like anomaly.
}
\label{Fig3}
\end{center}
\end{figure}
%
%
%
%
%
\begin{figure}
\begin{center}
\includegraphics[height=140mm,clip]{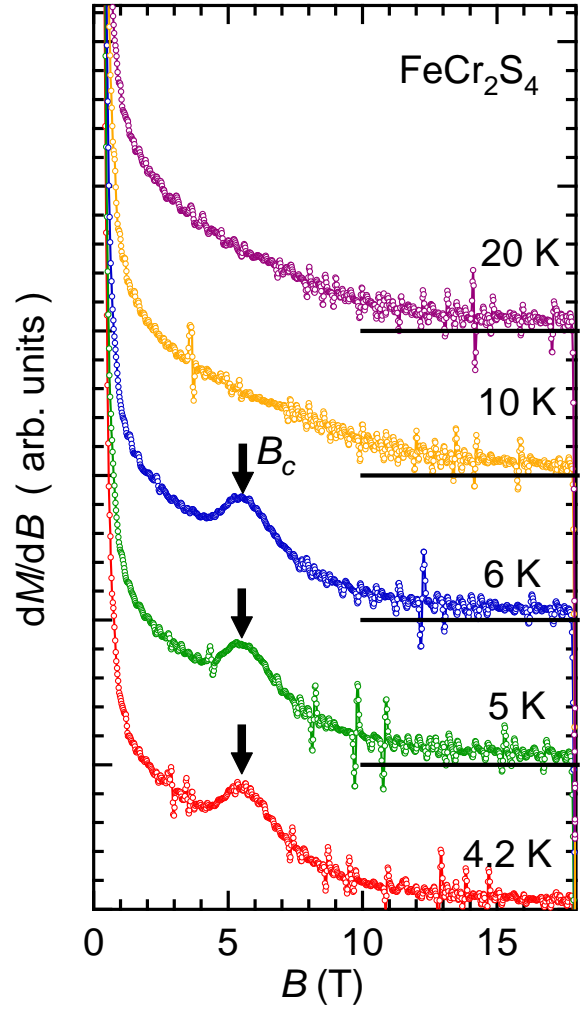}
\caption{
The magnetic field derivation of magnetization $dM(B)/dB$ of FeCr$_{2}$S$_{4}$ as the function of $B$ at the various temperature.
The arrows show the magnetic field position of the anomalies. Every curve is shifted to avoid overlap.
}
\label{Fig4}
\end{center}
\end{figure}
\begin{figure}
\begin{center}
\includegraphics[height=140mm,clip]{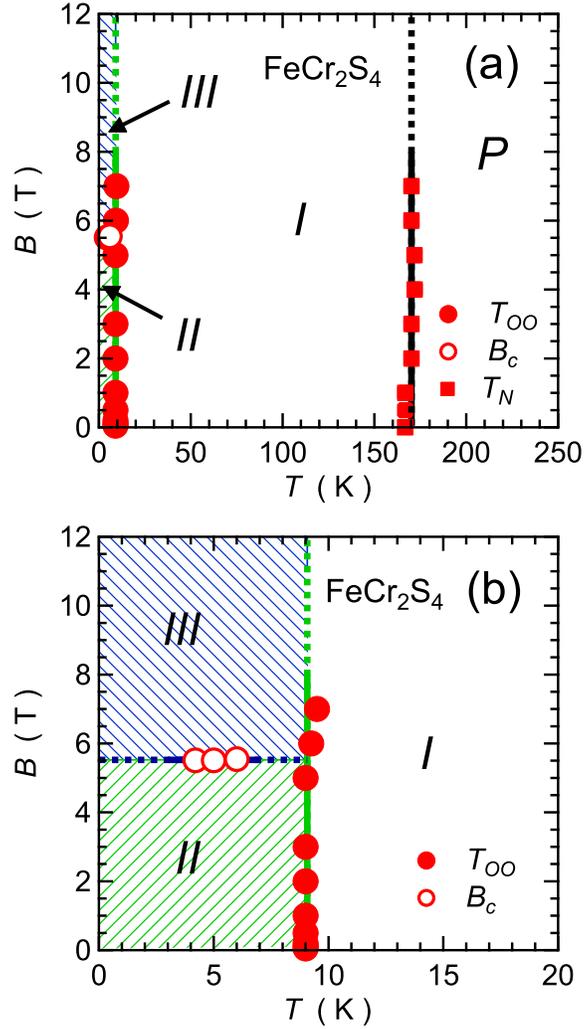}
\caption{
(a)The Magnetic field and temperature ($B-T$) phase diagram of FeCr$_{2}$S$_{4}$ determine by the results from $M(T)$ and  $M(B)$. (b) The plot with the expanded horizontal scale.  
The solid lines is guide to the eye.
The broken lines shows the expecting phase boundaries. }
\label{Fig5}
\end{center}
\end{figure}
%
%
%
%
%


\bibliographystyle{model1b-num-names}
\bibliography{<your-bib-database>}

\end{document}